\documentclass[]{aastex631}

\usepackage[T1]{fontenc}        

\received{...}
\revised{...}
\accepted{...}
\submitjournal{ApJ}

\shorttitle{Solar Flare Calcium Abundances}
\shortauthors{Sylwester et al.}
\graphicspath{{./}{figures/}}

\begin{document}


\title{New Solar Flare Calcium Abundances with no Surprises: Results from the SMM Bent Crystal Spectrometer}

\correspondingauthor{K. J. H. Phillips}
\email{kennethjhphillips@yahoo.com}

\author[0000-0002-8060-0043]{J. Sylwester}
\affiliation{Space Research Centre, Polish Academy of Sciences (CBK PAN), Warsaw, Bartycka 18A, Poland}
\email{js@cbk.pan.wroc.pl}

\author[0000-0001-8428-4626]{B. Sylwester}
\affiliation{Space Research Centre, Polish Academy of Sciences (CBK PAN), Warsaw, Bartycka 18A, Poland}
\email{bs@cbk.pan.wroc.pl}

\author[0000-0002-3790-990X]{K. J. H. Phillips}
\affiliation{Scientific Associate, Earth Sciences Department, Natural History Museum, Cromwell Road, London SW7 5BD, UK}
\email{kennethjhphillips@yahoo.com}

\author[0000-0002-5299-5404]{A. K\k{e}pa}
\affiliation{Space Research Centre, Polish Academy of Sciences (CBK PAN), Warsaw, Bartycka 18A, Poland}
\email{ak@cbk.pan.wroc.pl}

\begin{abstract}
The calcium abundance in flare plasmas is estimated using X-ray spectra from the Solar Maximum Mission Bent Crystal Spectrometer (BCS) during the decays of 194 flares (GOES classifications from B6.4 to X13) occurring between 1980 and 1989. Previous work by Sylwester et al. found that the abundance varied from flare to flare. That analysis is improved on here using updated instrument parameters and by including all calcium lines viewed by the BCS instead of only the resonance line, so greatly enhancing the photon count statistics. The abundance variations are confirmed with the average abundance, $A({\rm Ca})$ (expressed logarithmically with $A({\rm H}) = 12$), equal to $6.77 \pm 0.20$ for 194 flares (141 of which are new in this study). This range corresponds to factors of between 1.7 and 7.2 larger than the photospheric abundance and so our results are in line with a ``FIP" (first ionization potential) effect whereby low-FIP elements like Ca (FIP = 6.11~eV) have enhanced coronal abundances. The Ca flare abundance is uncorrelated with solar activity indices, but weak correlations are suggested with GOES flare class and duration (larger $A({\rm Ca})$ for smaller and shorter flares). The ponderomotive force theory of Laming explaining the FIP effect gives a range of parameters within which our estimates of $A({\rm Ca})$ agree with the theory. However, this then gives rise to disagreements with previous estimates of the flare silicon and sulfur abundances, although those of argon and iron are in good agreement. Small adjustments of the theory may thus be necessary.
\end{abstract}





\section{Introduction} \label{sec:intro}

Element abundances in both the quiet and active solar corona continue to be of great interest, with many observations of line emission made by instruments working in soft X-ray and extreme ultraviolet wavelengths adding considerable amounts of data to bear on the problem. These data have hitherto indicated that for the most part a ``FIP effect" operates, i.e., the abundances of elements differ from photospheric abundances depending on whether their first ionization potentials (FIP) is less  or more than about 10~eV. \cite{mey85} suggested that high-FIP elements were depleted with respect to photospheric abundances, but the later work of \cite{fel92b} based on ultraviolet and soft X-ray solar spectra instead indicated that low-FIP elements were enhanced in coronal structures and that the abundances of high-FIP elements were essentially photospheric. Results based on solar spectroscopy (e.g., \cite{vec81}) and work done more recently has revised these conclusions to some extent. Related to this are studies of averaged abundances in solar energetic particles and in the slow solar wind (e.g., \cite{rea14,rea18}), with similar conclusions although there are indications that the cross-over energy of the FIP may apparently differ from 10~eV depending on the element (a similar conclusion was also reached by \cite{den15}).

For soft X-ray flares, data are now available from high-resolution crystal spectrometers such as the RESIK (Rentgenovsky Spekrometr s Izognutymi Kristalami) instrument on the CORONAS-F spacecraft (wavelength range 3~--~6~\AA, operational 2001~--~2003: \cite{jsyl05,bsyl15}) and broad-band spectrometers with spectral resolution between about $100 - 200$~eV  \citep{nar14,cas15,kat20,vad21} and $0.6 - 1$~keV (at 6~keV) \citep{phiden12,den15}. The RESIK and broad-band spectrometers viewed the flare continuum, so absolute abundances can be inferred from line-to-continuum ratios. Spectral line features are identifiable from spectra observed with broad-band instruments as unresolved lines of abundant elements such as Si, S, Ar, Ca, and Fe. Of these elements Ne, Ar, and S have relatively high ($\gtrsim 10$~eV) FIPs and Ca and Fe are low-FIP elements. For solar X-ray flares, the estimated abundances from the works cited are enhanced for the low-FIP elements Ca and Fe over photospheric by factors of between 2 and 4, while Si (low-FIP) and S (FIP $= 10.4$~eV) have abundances near to or even lower than photospheric. Observations by RESIK indicate that Ar is close to proxies of the photospheric abundance \citep{bsyl15,lod08}. In the extreme ultraviolet, spectral and spatial flare observations include those made by the Extreme-ultraviolet Imaging Spectrometer (EIS) instrument on the Japanese Hinode spacecraft (e.g., \cite{bro15,dos16,dos19,bak19,to21}). These observations use the intensity ratio of a pair of Ca and Ar lines as a probe of the FIP effect, and show from maps of this ratio that the Ca line emission is enhanced in the main flare structures but in localised patches near sunspots the argon line emission is enhanced, giving rise to an inverse FIP effect \citep{dos19}.

The FIP effect has also been extensively investigated for stars with X-ray-emitting coronae using spectra from the Chandra mission with the Low Energy Transmission Grating (LETG) \citep{lam95,hue13a}. Recent findings indicate that for stars with later spectral types there is an inverse FIP effect that sharply increases towards spectral type M \citep{hue03,arg04,wood18}.

Although several ideas have been expressed over the years to explain the FIP or inverse FIP effect in coronal and in particular flare plasmas (e.g., \cite{hen98}), much attention in recent years has focused on that of \cite{lam04,lam21}, in which the ponderomotive force associated with Alfv\'{e}n or fast-mode waves transports ions of low-FIP elements to give rise to enhancements or depletions of elements.  A ponderomotive force arises in a region of inhomogeneous magnetic field, whose magnitude is given by the height gradient of the square of the electric field associated with the transporting waves divided by the square of the surrounding magnetic field. The exact region in the solar atmosphere where ions and any neutral atoms (which are unaffected by the ponderomotive force) is where the plasma $\beta$ (ratio of plasma to magnetic pressure) is of order one, corresponding to the upper chromosphere. At this location, low-FIP elements like Ca and Fe exist in a predominantly ionized state and high-FIP elements like Ar are largely neutral. A competition between downward-propagating Alfv\'{e}n waves from the corona and upward-propagating fast-mode waves from the lower atmosphere is thought to determine whether the ponderomotive force is directed upwards (for a FIP effect) or downwards (for an inverse FIP effect).

An extensive analysis of ionized calcium X-ray spectra at around 3.17~\AA\ from the Bent Crystal Spectrometer (BCS) on the NASA spacecraft  Solar Maximum Mission (SMM) by \cite{jsyl84,jsyl98} gave the surprising result that the abundance of calcium varied substantially from flare to flare. A broadly similar result was obtained with ionized calcium spectra from the U.S. Naval Research Laboratory's solar X-ray spectrometer SOLFLEX on the P78-1 spacecraft \citep{ste93}. The calcium lines were those of mainly He-like Ca (\ion{Ca}{19}) with \ion{Ca}{18} dielectronic satellites, and were observed by channel~1 of the BCS. The motivation in these analyses has been the fact that the BCS remains the instrument having the highest spectral resolution of any in the range $1.7 - 3.2$~\AA\ from the SMM era in the 1980s to the present. Also hundreds of medium-class flares and many major flares were recorded in this period which included the maximum of the flare-prolific solar cycle 21 through to the beginning of Cycle 22. Importantly, BCS channel~1 includes portions of the continuum emitted by flares which are uncontaminated by crystal fluorescence, as was established through detailed analysis by \cite{jsyl20}. This is probably true for the \ion{Ca}{19} channel of a similar but uncollimated crystal spectrometer on the Yohkoh spacecraft \cite{cul91}, but the lack of collimation leads to uncertainty in characterizing the continuum which arose from multiple sources on the Sun. The BCS continuum emission can be related to that in the two X-ray channels of the Geostationary Operational Environmental Satellites (GOES), specifically GOES~2, 5, and 8, which observed X-ray emission from the whole Sun during the SMM period and from which temperature information can be obtained from the ratio of emission in the two channels. The relation between GOES broad-band measurements and the BCS spectra will be given in work in preparation.

Here we report on a new analysis which has resulted in calcium abundance determinations for the decay phases of 194 flares observed over the lifetime of SMM, 1980 February to 1989 November. The present work improves on the \cite{jsyl98} analysis in three important respects. Firstly, account is now taken of non-uniformities in the cylindrical geometry of the bent diffracting crystal used in BCS channel~1 which are significant and affect the estimation of the emitting temperature in each spectrum through changed line intensities. Secondly, rather than measuring line-to-continuum ratios using only the \ion{Ca}{19} resonance (or $w$) line, as in our earlier work, we now use the entire calcium line spectrum viewed by BCS channel~1, including all the \ion{Ca}{19} lines and dielectronic satellites. Thirdly, we compare observed spectra with those calculated using the best available atomic data for the numerous dielectronic satellites present and the free--bound and free--free continua estimated using solar element abundances from the {\sc chianti} atomic code which was not available at the time of the work by \cite{jsyl98}. Our analysis covers activity cycles 21 and 22. We re-analyzed the 53 flares that were part of the \cite{jsyl98} analysis and included them in the flares of this analysis. A comparison of the Ca abundances from the \cite{jsyl98} analysis with the present one is given in Section~\ref{sec:meas_Ca_abunds}.

The total range of estimated abundances from our present work is a factor of more than two, much greater than the estimated uncertainties in individual determinations, so we conclude that substantial variations in the calcium abundance from flare to flare are indeed real, as was claimed by \cite{jsyl98}. We examine the relationships between determined abundances and other parameters, in particular flare class and duration for which there is a slight positive correlation. Finally, our results are discussed in terms of the Laming ponderomotive force theory.

\section{Selection of BCS Data} \label{sec:Data_selection}

The Solar Maximum Mission was launched on 1980 February~14 and operated successfully for a further nine months when the spacecraft's attitude control unit failed, leaving the spacecraft unable to point accurately at features on the Sun such as active regions or flares. The BCS was as a result unable to operate at all. A Space Shuttle Repair Mission (STS-41C) in 1984 April resulted in the replacement of the faulty attitude control unit and, very shortly after the SMM spacecraft was released from the Shuttle, operations recommenced, with the BCS able to observe flares as before.

The BCS consisted of eight channels, with channel~1 viewing the ionized calcium lines and with the remaining seven channels viewing lines of ionized Fe. The BCS was described in an early pre-launch paper by \cite{act80}; more detailed work on the calibration and refinement of BCS characteristics has since been done (\cite{rap17}; \cite{jsyl20}). Solar X-rays were incident via a coarse (6~arcminute square) collimator on to eight curved germanium crystals, with the Bragg-diffracted radiation detected by position-sensitive proportion counters, each with 256 spectral bins (occasionally a mode was used with double binning, giving a total of 128 bins). Read-out of each detector in pre-programmed integration times (typically a few seconds in order to record any rapid changes in flare emission) enabled complete spectral coverage over the range of each detector to be obtained instantaneously, without any scanning motion as is needed for flat crystal spectrometers.

As with our previous analyses \citep{jsyl98,rap17,jsyl20}, recourse was had to the SMM BCS data archive held on a NASA ftp site{\footnote{The site is ftp://umbra.nascom.nasa.gov. BCS data files are in the subdirectory pub/smm/xrp/data/bda.}}. In the work by \cite{jsyl98}, a total of 146 flares was selected over the SMM lifetime. Some of the data in the \cite{jsyl98} analysis are now unavailable in the data archive, but we were still able to select 194 flares, with 53 flares in common with the earlier analysis. The X-ray flare importance ranges from B6.4 to X13. From these 194 flares, more than 13000 individual spectra during a total of 2806 subintervals were selected.

Only spectra during the flare decay phases were selected. The durations of the subintervals were chosen to provide acceptable count statistics and generally only slowly varying temperature as determined from the emission ratio of the two GOES X-ray channels. Solar flare X-ray emission is far from being isothermal, with comparatively low-temperature line emission (e.g., \ion{Fe}{17}, $T \sim 1.5$~MK) being evident at the same time as line emission with very high temperatures; thus, for moderate to intense flares, He-like Fe (\ion{Fe}{25}, $T\gtrsim 12$~MK) line emission was commonly seen at times immediately following the flare impulsive phase. The plasma emitting these different line emissions is likely to be contained in different loops thermally insulated from each other. Confining the selection of BCS spectra to times a little later than the peak ensures that \ion{Fe}{25} line emission had diminished to almost zero  (as indicated by emission in the BCS channels viewing these lines). The next observable hottest line emission is in effect from He-like Ca (elements intermediate in atomic number, $Z=21-25$, have very low solar abundances and were not seen by the BCS). The \ion{Ca}{19} emission seen in channel~1 has dielectronic satellite lines from which (using the intensity ratio of satellites to the \ion{Ca}{19} $w$ line) a temperature is derived that is an average of this hottest part of the flaring plasma giving rise only to \ion{Ca}{19} emission with the \ion{Fe}{25} line emission having faded. The satellite-to-resonance line temperature is then assumed to be applicable to the line emission and the continuum seen in BCS channel~1, so relating the observed line-to-continuum ratio to the theoretical ratio.

Table~\ref{tab:list-of-flares} gives for each flare the date, peak time (UT) (cols. 2, 3) of the associated flare, NOAA active region number and heliographic coordinates (cols.~4, 5), the H$\alpha$ and GOES X-ray importance (cols. 6, 7), time range (UT) of the observations (cols.~8, 9), and the estimated mean calcium abundance from this analysis (col. 10; to be discussed later). For a few very intense flares the decay phase lasted more than one spacecraft orbit; for these flares BCS spectra were analyzed for each orbit and a calcium abundance derived. Column~11 of this table lists the number of subintervals (nI) for each flare, with nS the number of spectra accumulated by the BCS in this time. The table footnote explains the various remarks in col.~12.

\begin{deluxetable*}{rccrcrccccrc}
\tabletypesize{\small}
\tablecaption{SMM BCS flares in this analysis. \label{tab:list-of-flares} }
\tablewidth{0pt}
\tablehead{
\colhead{No.}  & \colhead{Flare Date} & \colhead{UT} & \colhead{AR} & \colhead{Location} & \colhead{H$\alpha$} & \colhead{X-ray} & \colhead{Start} & \colhead{End} & \colhead{$A({\rm Ca})$} & \colhead{nI/nS} & \colhead{Remarks$^a$} \\
&&&&& \colhead{class} & \colhead{class} \\
(1) & (2) & (3) & (4) & (5) & (6) & (7) & (8) & (9) & (10) & (11) & (12) \\}
\startdata
1&1980-Mar-29&09:20&2363&N21E39&SB&M1.0& 09:20:15&09:29:54&6.89 $\pm$ 0.15& 8/ 28& sl \\
2&1980-Apr-07&18:45&2372&N12W08&SB&M1.3& 18:46:03&18:59:55&6.75 $\pm$ 0.03&10/ 46& sl \\
3&1980-Apr-08&03:08&2372&N12W13&1B&M4.6& 03:10:00&03:22:43&6.72 $\pm$ 0.02&12/ 38& c \\
4&1980-Apr-10&09:22&2372&N12W42&1N&M4.1& 09:22:10&09:35:43&6.73 $\pm$ 0.02&15/ 45& c \\
5&1980-Apr-11&23:14&2372&N10W70&1N&M1.6& 23:15:08&23:29:52&6.81 $\pm$ 0.07&16/ 49 \\
6&1980-Apr-13&04:09&2372&N10W77&1F&M1.4& 04:12:08&04:37:17&6.74 $\pm$ 0.03&25/132& c \\
7&1980-Apr-22&05:23&2396&S12E28&SN&C8.7& 05:24:09&05:49:43&6.74 $\pm$ 0.04&17/ 85 \\
8&1980-Apr-30&20:26&2396&S13W90&SN&M2.2& 20:28:01&20:39:50&6.72 $\pm$ 0.03&10/ 39& c \\
9&1980-May-09&07:14&2418&S21W32&1B&M7.2& 07:15:15&07:22:50&6.71 $\pm$ 0.02& 8/ 25&nG \\
10&1980-May-21&21:07&2456&S14W15&2B&X1.5& 21:08:07&21:29:54&6.69 $\pm$ 0.01&23/116& c sl \\
\\
\enddata
\tablecomments{Table 1 is published in its entirety in machine-readable format. The first ten data entries are shown here to indicate the table's form and content.}
\tablenotetext{a}{Remarks: Abundances $A({\rm Ca})$ are expressed logarithmically with $A({\rm H})=12$. Stated uncertainties include any time variations; nI/nS  (col.~11): nI = number of subintervals in flare duration, nS = number of BCS spectra. Remarks (col.~12): c Flare also analyzed by \cite{jsyl98}; nG GOES data not available; Gs GOES flux saturated; db double spectral binning; dbnw double spectral binning except w-line region; sl simple GOES lightcurve.}
\end{deluxetable*}

\section{Analysis Methods} \label{sec:anal_methods}

\begin{figure}
\gridline{\fig{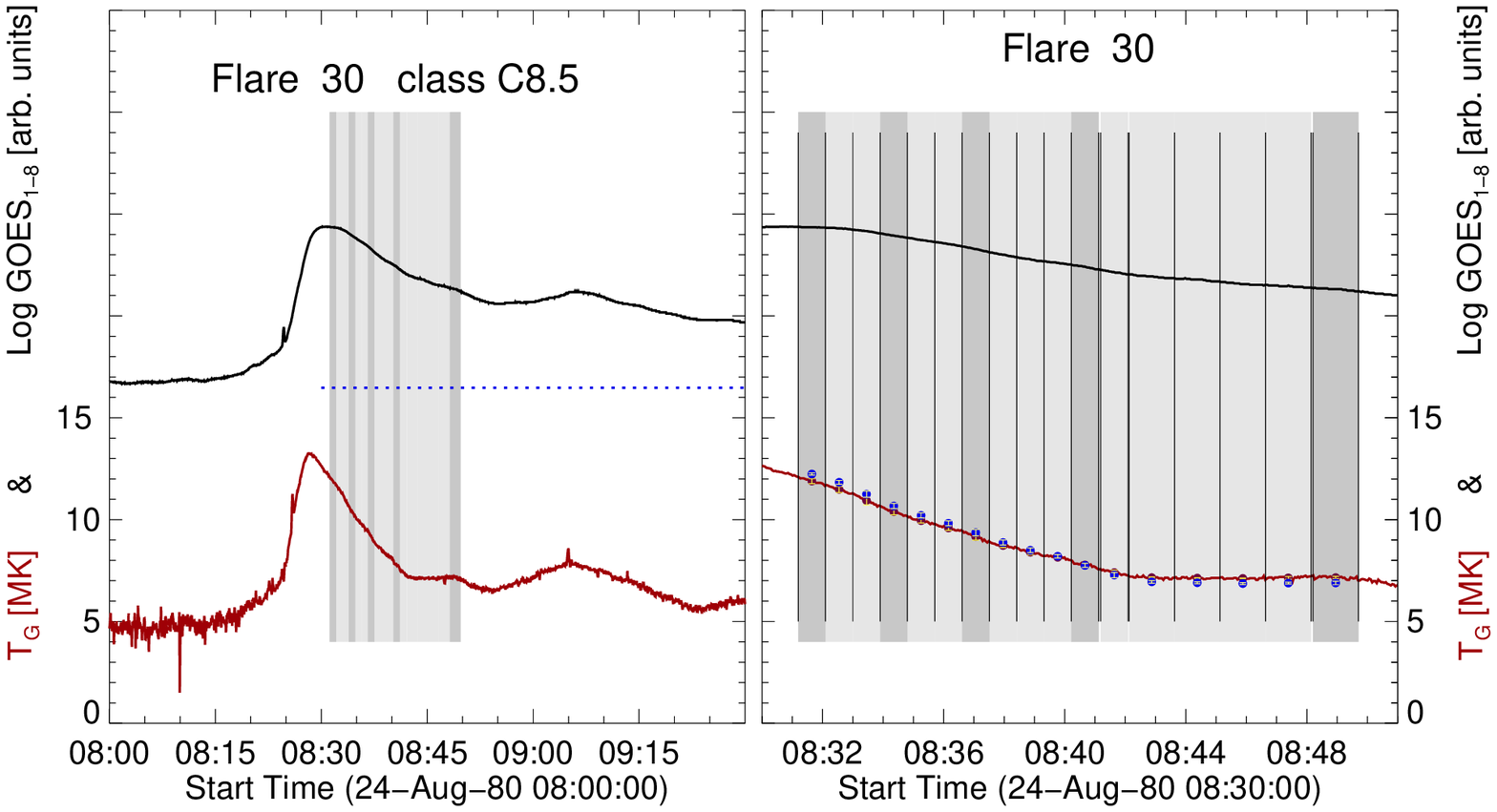}{0.45\textwidth}{(a)}
  \fig{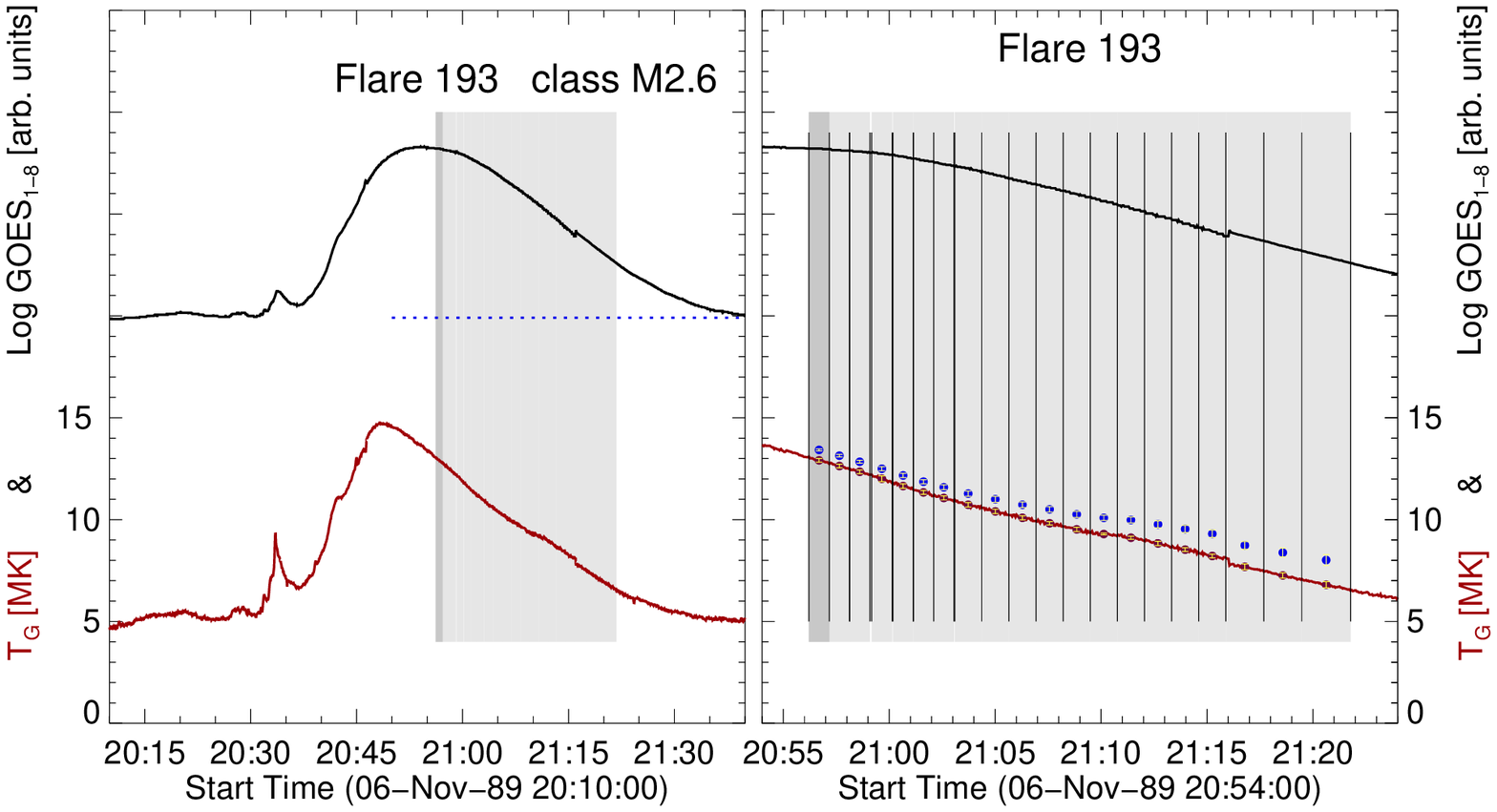}{0.45\textwidth}{(b)} }
\caption{Light curves (left panels of $a$, $b$) and time subintervals (right panels of $a$, $b$) for flares on 1980 August~24 and 1989 November~6 (flares 30 and 193 in Table~\ref{tab:list-of-flares}). Between 3 and 5 averaged spectra were included over the subintervals indicated by the gray-colored strips for flare 30 and between 3 and 9 spectra for the gray-colored strips for flare 193. Averaged spectra in the strips colored dark gray for flare 30 are shown in Figure~\ref{fig:fig5}.
\label{fig:fig1}}
\end{figure}

We illustrate the analysis methods used here with spectra during the C8.5 flare on 1980 August~24 (number~30 in Table~\ref{tab:list-of-flares}) and the M2.6 flare on 1989 November~6 (number~193). These flares chosen from near the beginning and end of spacecraft operations exemplify how BCS spectra changed over the SMM lifetime. Figure~\ref{fig:fig1} (left panels for each flare) shows the GOES X-ray light curve in its 1~--~8~\AA\ channel and the temperature $T_{\rm G}$ derived from the emission ratio for the two GOES channels. Individual spectra during the decays of each of these flares were analyzed in subintervals, with the vertical lines in each of the right-hand panels $a$ and $b$ indicating the subinterval time ranges.

%

\begin{figure}
\centerline{\includegraphics[width=0.75\textwidth,clip=]{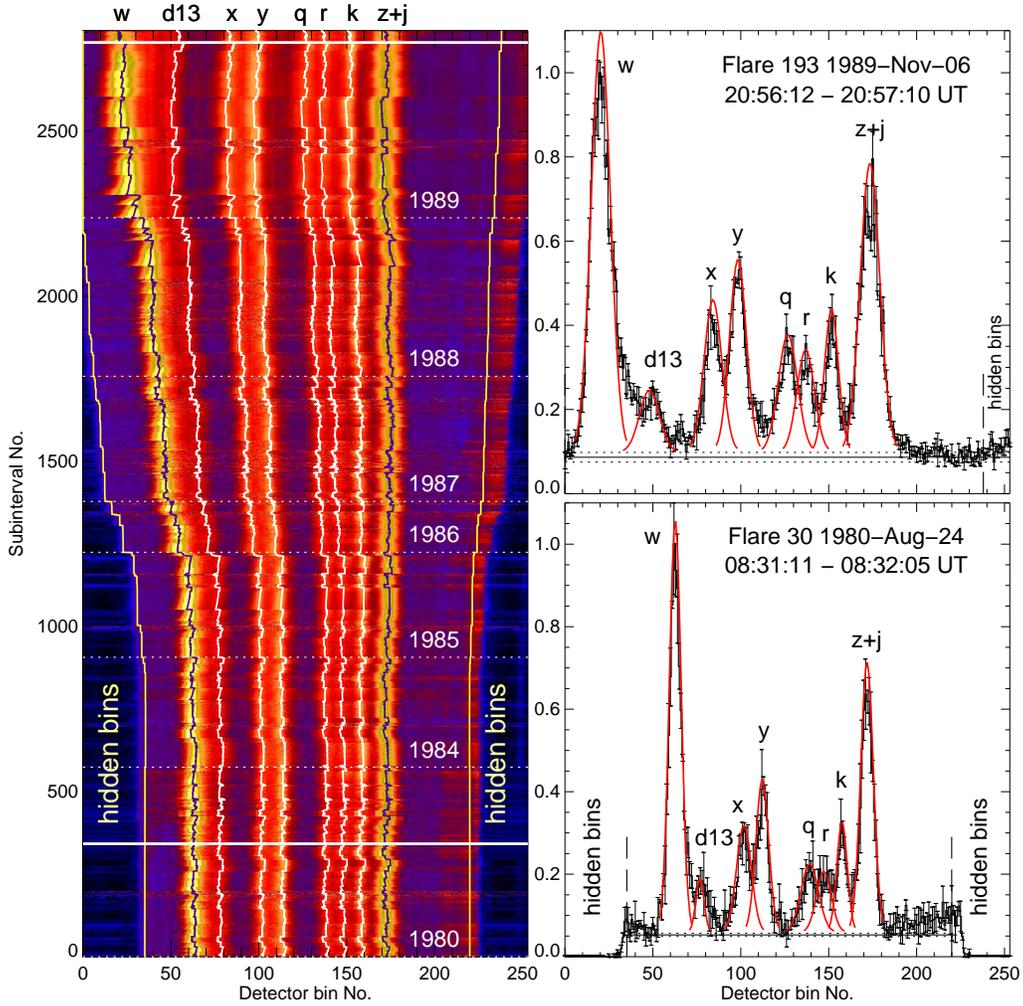}}
\caption{Left panel: stacked BCS channel 1 spectra for subintervals during all flares listed in Table~\ref{tab:list-of-flares}. Estimated positions of ``hidden'' (non-solar) bins at the lower and upper ends of the spectra are shown, based on \cite{rap17}. Time proceeds upwards, with dotted horizontal lines marking year boundaries, and the thick horizontal lines marking flares 30 and 193. Right panel: BCS channel~1 spectra during the first subintervals of flares 193 (upper) and 30 (lower), with the eight principal line features ($w$, $d13$, $x$, $y$, $q$, $r$, $k$, and $z+j$) labeled and estimated continuum level. The spectra are plotted against bin number rather than wavelength to illustrate the change in dispersion (wavelength interval per unit bin range) between 1980 and 1989. The locations of the BCS hidden  bins are indicated by dashed lines. Error bars represent statistical uncertainties in photon counts in each bin.
\label{fig:fig2}}
\end{figure}

In Figure~\ref{fig:fig2} (left panel), BCS channel~1 spectra for the entire set of 2806 subintervals are shown in a stack form, with time progressing upwards. The narrow light-colored regions are spectral lines due to He-like calcium (\ion{Ca}{19}) and dielectronic \ion{Ca}{18} satellite lines. Identifications are indicated at the top of the stack (full details including wavelengths and line identifications were given in Table~1 of \cite{jsyl20}). As is evident, the dispersion (wavelength range in m\AA\ per spectral bin) decreased over time during the spacecraft lifetime, an effect due to the detector electronics that was discussed by \cite{rap17} and \cite{jsyl20}. There are ``hidden'' or non-solar bins at either end of the spectral range whose original purpose was to monitor the cosmic ray background rate. These are well-defined in the 1980 period but after 1985, approximately one year after the Space Shuttle Repair Mission, the hidden bins become progressively narrower, with the short-wavelength range becoming obliterated by the end of 1988. The bin positions of the hidden bins were determined by the spectral dispersion which in turn was determined by the bin positions of the prominent spectral line $w$ (wavelength for an unshifted source 3.17735~\AA: \cite{phi18}) and the blended $z + j$ line feature (3.21095~\AA). In the right panels of Figure~\ref{fig:fig2} are two example spectra for the first subintervals of flares 30 and 193, with integration times of less than a minute, showing photon counts per bin normalized to the peak of \ion{Ca}{19} line $w$ with error bars indicating statistical uncertainties. The abscissa is in bin number rather than absolute wavelength in \AA\ to show the decreasing dispersion with time. A base level for each spectrum, estimated from the solar bins beyond the $w$ and $z + j$ line features (flare 30) and from the solar bins beyond the $z + j$ line feature alone (flare 193), is as mentioned due to continuum emission from each flare, with negligible crystal fluorescence.

%

\begin{figure}
\centerline{\includegraphics[width=0.9\textwidth,clip=]{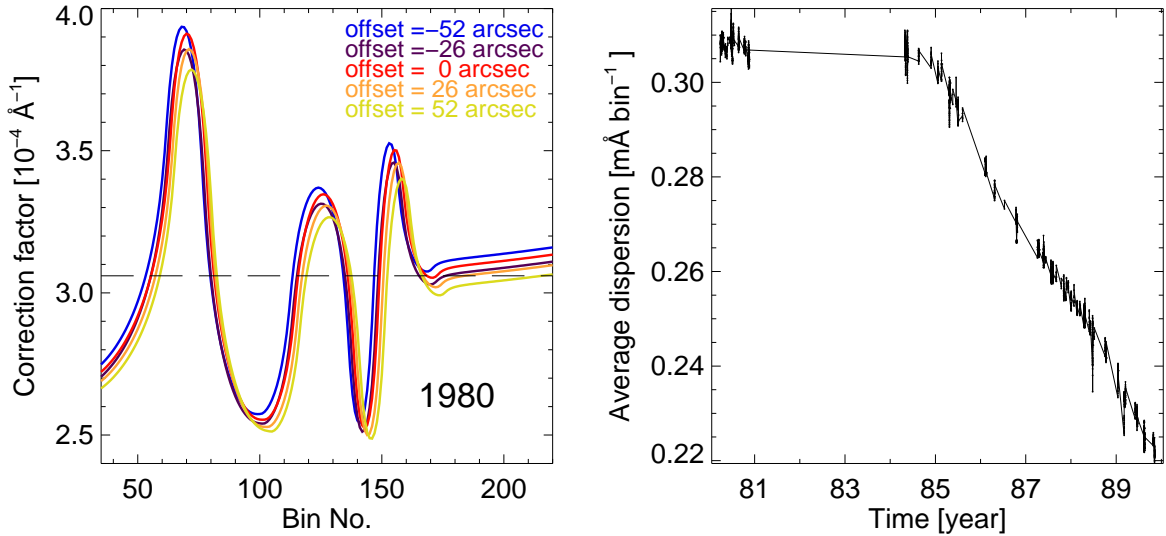}}
\caption{Left panel: Correction factors (in $10^{-4}$ \AA$^{-1}$) as a function of BCS channel~1 bin number for various offsets (indicated by different colors) from the BCS boresight during  the 1980 period. These factors were applied to spectral bins to obtain reduced channel~1 spectra. The dashed horizontal line represents the value of the correction factor for an ideal crystal bent to a perfect cylindrical surface. Right panel: Dispersion (m\AA\ bin$^{-1}$) of BCS channel~1 over the SMM spacecraft lifetime. No observations were taken between 1980 November and 1984 April when precise spacecraft attitude was lost.
\label{fig:fig3}}
\end{figure}

\begin{figure}
\centerline{\includegraphics[width=0.75\textwidth,clip=]{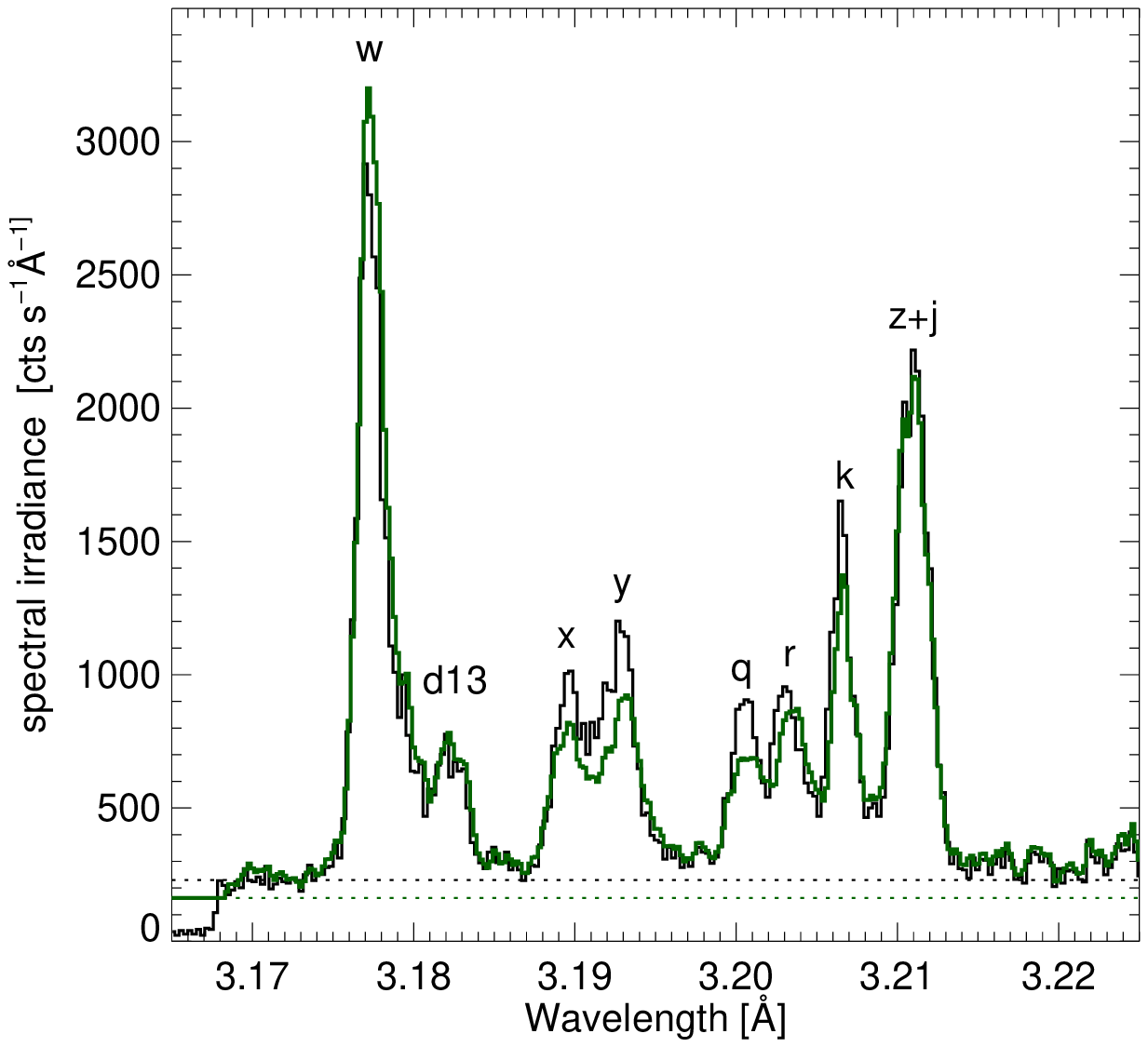}}
\caption{BCS channel~1 spectra for the decay of flare 113 on 1987 April 17 at 02:36:05 UT) with (green histogram) and without (black histogram) correction factors for the non-uniformities in the diffracting crystal.
\label{fig:fig4}}
\end{figure}

Figure~\ref{fig:fig3} (left panel) shows correction factors due to crystal curvature which were applied to BCS spectra to obtain reduced spectra. These factors, previously discussed by \cite{jsyl20} in re-examining BCS instrument parameters, express enhancements to portions of BCS spectra through non-uniformities in the cylindrical geometry of the channel~1 bent crystal; these non-uniformities in turn give rise to changes in crystal reflectivity and so observed line intensities. Their application to a channel~1 spectrum (flare 113, 1987 April~17) is illustrated in Figure~\ref{fig:fig4} where the black histogram is the uncorrected spectrum and the green histogram is with the correction factors. This particular spectrum shows how a long-standing discrepancy in the \ion{Ca}{19} $x$ and $y$ line intensities with atomic theory (which predicted the $x:y$ intensity ratio to be nearly one; \cite{bel82b}) was resolved. For the present work, the correction factors are relevant for the \ion{Ca}{19} line $w$ and the well-resolved \ion{Ca}{18} satellite line $k$ since their intensity ratio gives the emitting temperature for each spectrum. From Figure~\ref{fig:fig4}, the $k/w$ ratio is 0.52 for the black histogram, but 0.40 for the green, leading to temperatures of 8.7~MK and 9.8~MK respectively. This $\sim 1$~MK difference is typical of most spectra we analyzed.

Figure~\ref{fig:fig3} (right panel) shows the time variation of the channel~1 dispersion (in m\AA\ bin$^{-1}$) as measured by the difference in bin positions of the \ion{Ca}{19} $w$ line and the $z + j$ line blend over the BCS lifetime, and shows what is already evident in Figure~\ref{fig:fig2} (left panel). As can be seen, the dispersion remained approximately constant over the 1980 period and for several months after the Shuttle Repair Mission in 1984 April but then declined linearly with time. These dispersion changes do not in principle affect the present analysis since the calcium abundances are determined from the ratio of the total line emission to the continuum, both of which are equally affected by the dispersion changes. Towards the end of the mission lifetime, particularly for flares in 1989, the absence of the observable continuum on the short-ward side of line $w$ makes the continuum level rather more difficult to determine, although using comparison with calculated spectra the continuum level can be found from that at longer wavelengths.

%

%

\begin{figure}
\centerline{\includegraphics[width=0.75\textwidth,clip=]{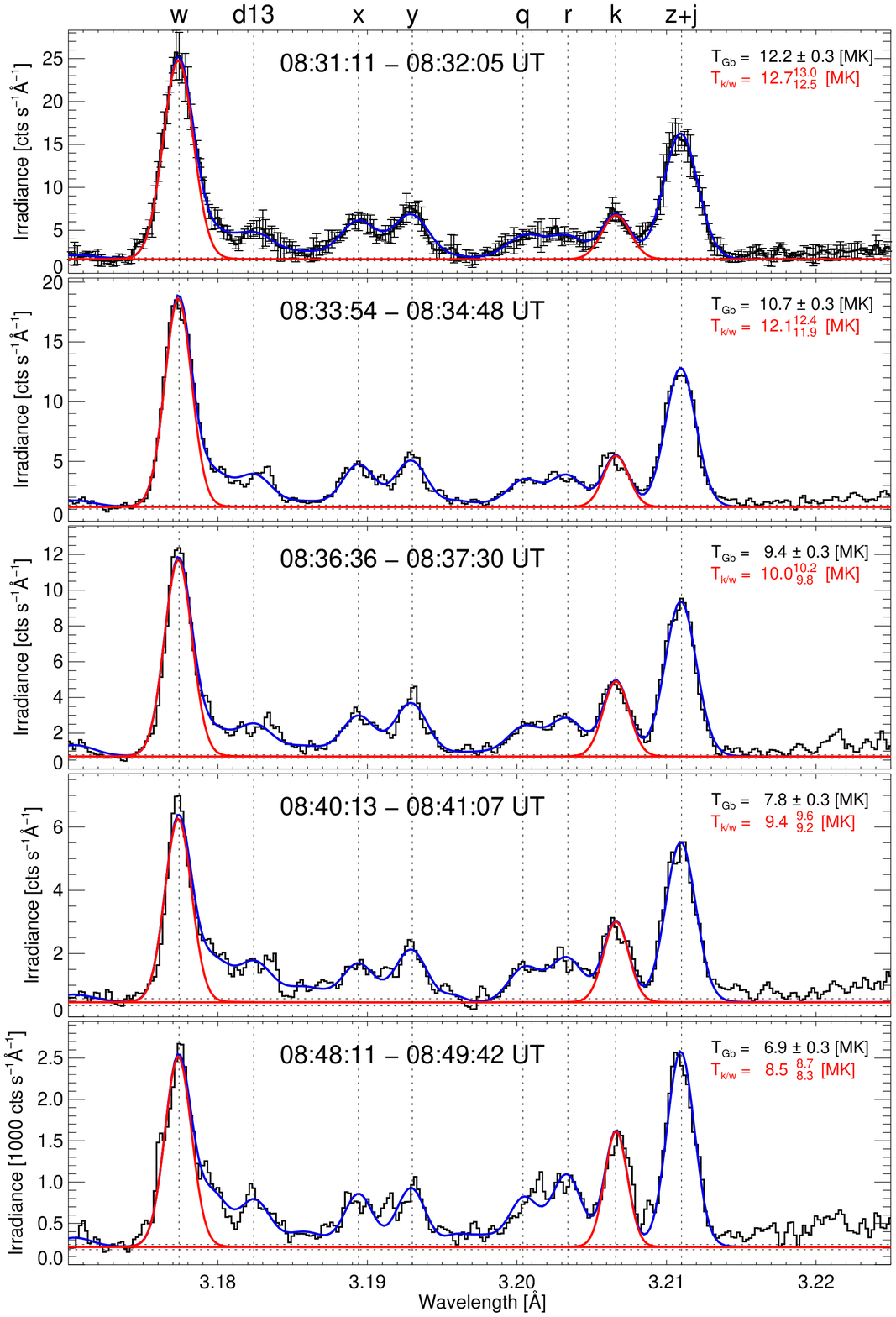}}
\caption{Five BCS channel 1 reduced spectra during the decay of flare 30 on 1980 August~24 (earliest times at the top). Spectral irradiance is in photon counts s$^{-1}$ \AA$^{-1}$. Vertical dashed lines indicate the eight principal spectral line features (Figure~\ref{fig:fig2}) and the sum of Voigt fits shown as the blue curve. The total observed flux above the continuum is used to calculate the line-to-continuum ratio from which the calcium abundance is obtained. The temperature $T_{\rm k/w}$ is determined from areas under the Voigt profiles to the $w$ and $k$ lines, shown in red. The nearly horizontal red line is the estimated continuum level (small error bars are uncertainties). The temperatures from GOES ($T_{\rm Gb}$) and from the line ratio $k/w$ ($T_{\rm k/w}$) are indicated for each spectrum. Error bars (in the top panel) are statistical uncertainties in photon counts. These five spectra correspond to the intervals colored dark gray in Figure~\ref{fig:fig1} ($a$, right panel).
\label{fig:fig5}}
\end{figure}

In Figure~\ref{fig:fig5} five spectra from the 1980 August~24 flare are illustrated with principal line features, marked as in Figure~\ref{fig:fig2}. The sum of the Voigt fits to these line features is shown as the blue curve in each spectrum, with red curves for the $w$ and $k$ line features separately. The temperature of the flare plasma for each spectrum was derived from the ratio of emission in the two GOES channels but with a pre-flare or post-flare background subtracted ($T_{\rm Gb}$ in the figure), and also from the $k/w$ line; this temperature is denoted by $T_{\rm k/w}$. As can be seen for each of the five spectra, the two temperatures are similar, but generally $T_{\rm k/w}$ is the larger of the two. The contribution to the BCS calcium spectra from the host non-flaring region was always negligible so the subtraction of a pre-flare level for the channel~1 spectra was unnecessary.

%

\begin{figure}
\centerline{\includegraphics[width=0.75\textwidth,clip=]{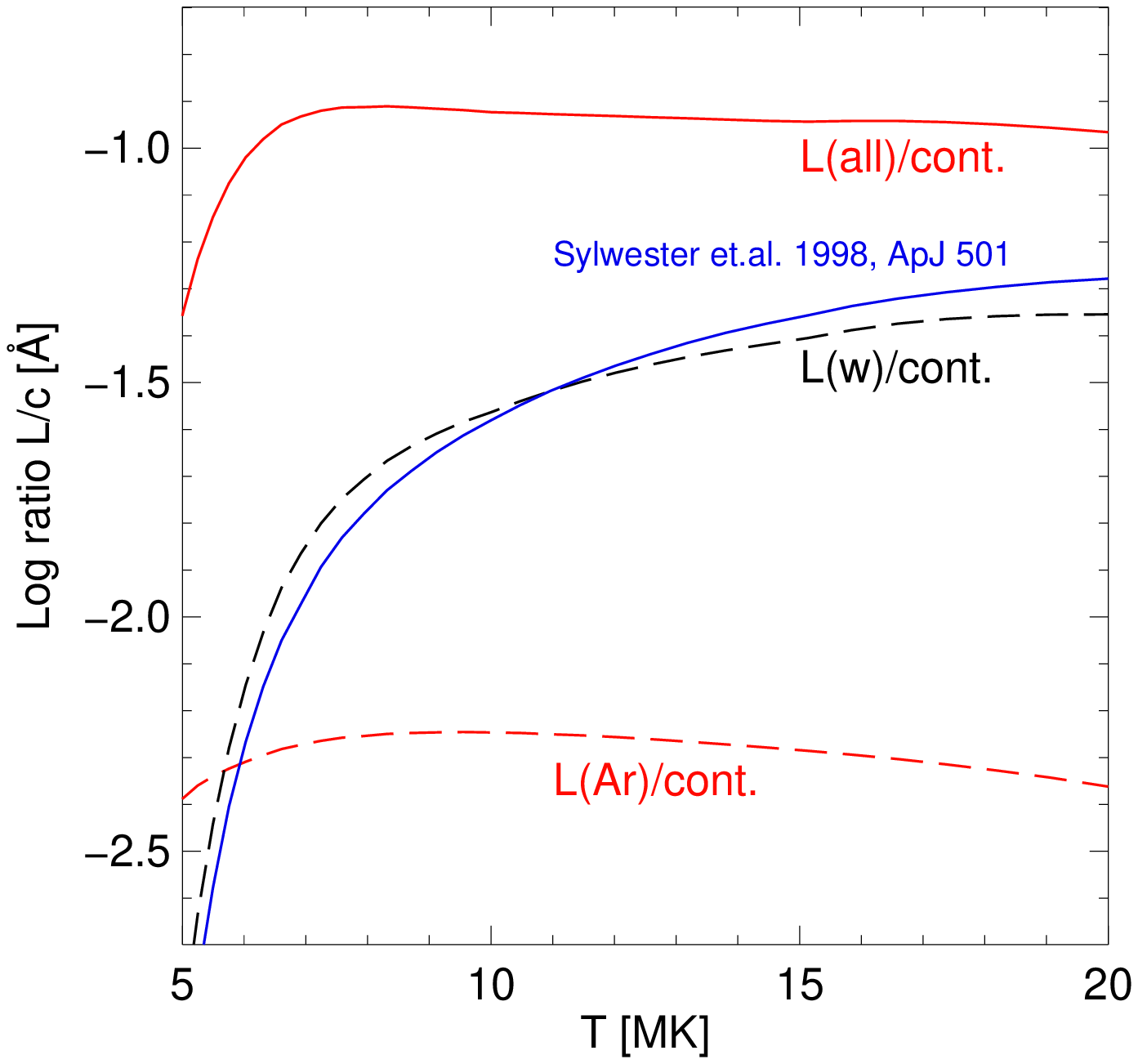}}
\caption{Line-to-continuum (at 3.1723\,\AA) ratios plotted logarithmically against temperature (MK) for all lines in BCS channel~1 (solid red line, this analysis), all argon lines in the channel~1 range (dashed red line) and the line $w$-to-continuum empirically determined by \cite{jsyl98} (blue curve) and present theory (dashed black line). See text for the assumed Ca and Ar abundances and temperature dependences.
\label{fig:fig6}}
\end{figure}

In the analysis of \cite{jsyl98}, the ratio of emission in line $w$ only to neighboring portions of BCS channel~1 spectra taken to be continuum was used to derive the calcium abundance during flare decays; the electron temperature was estimated as is done here from the $k/w$ line ratio. For the present analysis, the ratio of the total emission in the entire calcium line spectrum in channel~1 to the continuum is used to estimate the calcium abundance. Figure~\ref{fig:fig6} shows on a logarithmic scale the dependence of the line-to-continuum ratios, with the original \cite{jsyl98} line~$w$-to-continuum ratio (blue curve) and the present ratio using all lines in the channel~1 range (red line marked $L({\rm all})$/cont.). As can be seen, including all lines greatly improves the statistical quality of the observed ratio as the total line emission is a factor of between 2.5 and 4 larger (depending on temperature) than that of the $w$ line alone. An additional advantage is that the temperature dependence is flatter for the $L({\rm all})$/cont. curve, so the Ca abundance is less dependent on the derived temperature and on any multi-temperature effects. The total line emission due to both \ion{Ca}{19} and \ion{Ca}{18} satellites was calculated by using a grid of spectra over the wider range seen by the DIOGENESS instrument on the CORONAS-F spacecraft \citep{phi18}. A small contribution to the spectrum is made by line emission due to ionized argon, in particular the \ion{Ar}{17} $w4$ line (3.1997~\AA: transition $1s^2 - 1s4p$) which blends with the \ion{Ca}{18} $q$ line and a weak \ion{Ar}{16} dielectronic satellite (labeled A6 in \cite{phi18}) on the short-wavelength side of the $w$ line. The contribution of these argon lines was allowed for, although as Figure~\ref{fig:fig6} indicates the contribution is only a few percent for $T \gtrsim 7$~MK, the range of measurable temperatures. The curves in this figure are based on a calcium abundance $A({\rm Ca}) = 6.76$ (average of values found by \cite{jsyl98}) and $A({\rm Ar}) = 6.47 \pm 0.08$ (\cite{bsyl15}, based on RESIK flare spectra), or a Ar/Ca abundance ratio  of 0.51. Included in Figure~\ref{fig:fig6} is the line~$w$-to-continuum curve ($L(w)$/cont.) as would be estimated from this analysis; the shape of the \cite{jsyl98} curve was determined empirically, so explaining the slight difference from the present $L(w)$/cont. curve.

Taking the line $w$-to-continuum ratio and that of all emission lines in the channel~1 range to the continuum should give similar values of $A({\rm Ca})$, and this was checked to be the case. From Figure~\ref{fig:fig4}, the value of log~(line $w$/cont.) is estimated to be -1.412, giving $A({\rm Ca}) = 6.927$, while log~(all line emission/cont.) is -0.758, giving the very similar value $A({\rm Ca}) = 6.930$.

The continuum in the BCS channel~1 (3.171~--~3.226~\AA) range is made up of free--free and free--bound emission in comparable amounts, with much smaller amounts due to two-photon emission. The free--bound emission in this range is formed by several elements, so an abundance set must be chosen for an accurate assessment of its amount. The analysis of \cite{jsyl98} used the continuum as calculated by \cite{mew86} with cosmic abundances, but several sets of solar abundances are now available. Early estimates of coronal abundances such as those of \cite{fel92b} were based on a variety of sources. A more recent review was given by \cite{sch12}. Since these works, several abundance determinations have been made using X-ray spectroscopy, in particular the RESIK crystal spectrometer for which instrumentally formed fluorescence is either minimal or accurately assessed \citep{bsyl15}, so enabling line-to-continuum ratios to be derived. Observations have also been made with broad-band spectrometers, having energy resolution allowing only groups of lines to be distinguished rather than detailed line structure but these instruments are able to observe the solar flare continuum. They include the Reuven Ramaty High-Energy Solar Spectroscopic Imager (RHESSI) (germanium detectors observing the Fe lines at 6.7~keV with 1-keV resolution); the MErcury Surface, Space ENvironment, GEochemistry, and Ranging (MESSENGER) mission (solid state X-ray detectors with 0.6~keV resolution); and the Suzaku X-ray Imaging Spectrometer (XIS) (X-ray-sensitive CCD cameras with 180~eV resolution, observing large flares using Earth albedo measurements). The two Indian lunar missions, Chandrayaan-1 and Chandrayaan-2, carried Solar X-ray Monitor (XSM) payloads, with resolution of 180~eV at 5.9~keV, and observed several rather weak flares \citep{nar14,mon21} as well as the quiet solar minimum Sun. In Table~\ref{tab:abundances} we give an abbreviated list of these recently determined solar flare abundances (expressed on a logarithmic scale with H~$=12$) for major elements involved in the emission of free--bound continuum. We include results from the OSO-8 graphite crystal spectrometer \citep{vec81} since with negligible fluorescence they are likely to be as reliable as those from RESIK and BCS channel~1. The table also includes the measurements for the single GOES class~C flare reported by \cite{nar14} as, out of the several weak ones they observed, this is more representative of the BCS flares included here (Table~\ref{tab:list-of-flares}). Table~\ref{tab:abundances} also shows abundances determined from an average of solar energetic particle (SEP) events compiled by \cite{rea14}, and photospheric abundances from spectroscopy or proxies (for Ne and Ar).

To calculate the free--free and free--bound continua here, we adopted an abundance set that is based on the photospheric abundances for O and Ne, on the assumption that coronal abundances are nearly equal to photospheric values for these high-FIP elements; the SEP value for Mg; RESIK values (flare decays) for Si, S, and Ar; the SEP value and the averaged Ca abundance from \cite{jsyl98}; and the RHESSI value of \cite{phiden12} for Fe. The continua were calculated with {\sc chianti} routines \citep{delz15}. These indicate that major contributions to the free--bound continuum are made by (in decreasing order) Fe, Mg, Si, O, Ne, Ca, C, and N for temperatures here and the BCS channel~1 wavelength range. Calcium makes only a very small contribution, so there is no danger of a circular argument in determining $A({\rm Ca})$ from line-to-continuum ratios when the continuum itself depends on the calcium abundance. We found that the total continuum from the Table~\ref{tab:abundances} abundances was only slightly (8\% at 15~MK) lower than that from the \cite{sch12} abundances.

\begin{deluxetable*}{ccccccccc}   
\tabletypesize{\scriptsize}
\tablecaption{Selected Element Abundance Sets and Those Adopted Here$^a$ \label{tab:abundances} }
\tablewidth{0pt}
\tablehead{
\colhead{O} & \colhead{Ne} & \colhead{Mg} & \colhead{Si} & \colhead{S} & \colhead{Ar} & \colhead{Ca} & \colhead{Fe} & \colhead{Element}  \\
13.6 & 21.6 & 7.6 & 8.2 & 10.4 & 15.8 & 6.1 & 7.9 & (First ionization potential in eV) \\
&&&&&&&& \colhead{Reference}}
\startdata
8.89 & 8.08 & 8.15     & 8.10      & 7.27      & 6.58       & 6.93       & 8.10       & \cite{fel92b} (Cor. abund. review) \\
8.61 & 7.90 & 7.87     & 7.86      & 7.23      & 6.35       & 6.64       & 7.85       & \cite{sch12} (Cor. abund. review) \\
\\
     &      &          & 7.7(.25) & 6.9(.2)  & 6.38(.2)     & 6.51(.2)   &            & \cite{vec81} (Cor. flares, OSO-8) \\
     &      &          &           &           &            &            & 7.91 (.1)  & \cite{phiden12} (Cor. flares, RHESSI) \\
     &      &          & 7.47(.1)  & 6.94(.09) &            & 6.58(.18)  & 7.99(.09) &\cite{nar14} (C class flare, Chandrayaan-1)\\
     &      &          & 7.53(.08) & 6.91(.07) & 6.47(.08)  &            &            &\cite{bsyl15} (Cor. flares, RESIK) \\
     &      &          & 7.7(.2)   & 7.2(.2)   & 6.8(.2)    & 6.91(.09)  & 7.72(.11) & \cite{den15} (Cor. flares, MESSENGER)\\
     &      &          & 7.36      & 6.61      & 6.35       & 6.62       &            & \cite{kat20} (Coronal flares, Suzaku)$^b$\\
\\
8.71(.01) & 7.91(.02) & 7.96(.01) & 7.89(.1) & 7.11(.03) & 6.34(.04) & 6.75(.04) & 7.82(.03) & \cite{rea14}, SEP events$^c$ \\
\\
\\
8.69(.05)& 7.87 (.10)& 7.55(.02) & 7.54(.02) & 7.19(.04) &            & 6.34(.03)  & 7.47(.03) & \cite{lod03} (Photosph./meteoritic, review) \\
     &      &          &          &            & 6.50(.10)  &            &            & \cite{lod08} (Various non-solar sources) \\
8.71(.04)&8.15(.10)&7.56(.05)& 7.51(.03) & 7.15(.03) & 6.50(.10) & 6.32(.03) & 7.48(.04) & \cite{lod21} (Photospheric) \\
8.69(.04)&8.06(.05)&7.55(.03)& 7.51(.03) & 7.12(.03) & 6.38(.10) & 6.30(.03)& 7.46(.04) & \cite{asp21} (Photospheric) \\
\\
{\bf 8.7}   & {\bf 7.9} & {\bf 7.9}     & {\bf 7.53} & {\bf 6.91}& {\bf 6.47}& {\bf 6.76}       &  {\bf 7.91}       & {\bf Coronal abundances adopted here.} \\
\\
\enddata
\tablenotetext{a} {Elements important for free--bound emission (Ar and Ca included also). Abundances on a logarithmic scale with H=12 (uncertainties in parentheses).}
\tablenotetext{b} {Averaged values over several flares relative to \cite{lod03}.}
\tablenotetext{c} {Average of 54 SEP events: abundances expressed with reference to Si but converted here to logarithmic scale with H=12.}
\end{deluxetable*}

\begin{figure}
\centerline{\includegraphics[width=0.75\textwidth,clip=]{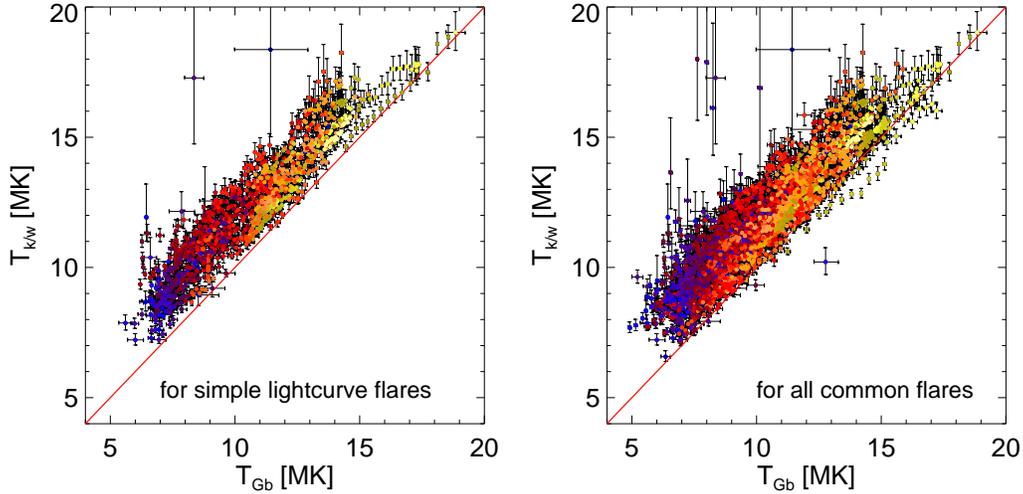}}
\caption{Temperature $T_{\rm k/w}$ for subintervals with simple GOES light curves (left panel) and with more complex GOES light curves  (right panel) plotted against the GOES temperature ($T_{\rm Gb}$). The colors of points relate to flare class from blue for B-class flares through red (M-class) to magenta (X-class) and yellow (X10 class). Of the 2806 subintervals, some 2718 had corresponding GOES light curves available, for which 1148 had simple light curves from which pre- or post-flare backgrounds could be easily subtracted.
\label{fig:fig7} }
\end{figure}

The analysis of DIOGENESS \ion{Ca}{19} flare spectra \citep{phi18} mentioned earlier showed that the temperature $T_{\rm k/w}$ was nearly equal to the temperature derived from the emission ratio of the two channels of GOES in the temperature range $12 - 22$~MK. As indicated earlier, because there is generally some contribution to the GOES X-ray emission from elsewhere on the Sun, a pre-flare or post-flare background emission should first be subtracted. This was readily done for flares with simple light curves, but otherwise pre-flare and post-flare X-ray emission levels were frequently more difficult to define. To investigate this, $T_{\rm k/w}$ was plotted against the background-subtracted GOES temperature $T_{\rm Gb}$ (Figure~\ref{fig:fig7}) for all the subintervals in both flares with simple light curves (left panel) and all flares (simple and non-simple flares: right panel). Subintervals with well-determined temperatures, indicated by small error bars, show a tight correlation, with $T_{\rm k/w}$ approximately 2~MK higher than $T_{\rm Gb}$ for simple flares. (This temperature difference was noted for the spectra in Figure~\ref{fig:fig5}.) The correlation is less clear when all flares are included, in particular those with pre-flare or post-flare levels that were more difficult to define.

%

The possibility of BCS detector saturation is relevant for very strong flares, of GOES class X and sometimes class M. The saturation reduces the amount of line emission, as would be expected, but a detailed analysis showed that the continuum emission was reduced by the same amount, so the measured line-to-continuum ratios were unaffected. The derived calcium abundances are thus still valid. This point will be dealt with in more detail in work in preparation.

\section{Measured Calcium Abundances}\label{sec:meas_Ca_abunds}

\begin{figure*}
\centerline{\includegraphics[width=0.75\textwidth,clip=]{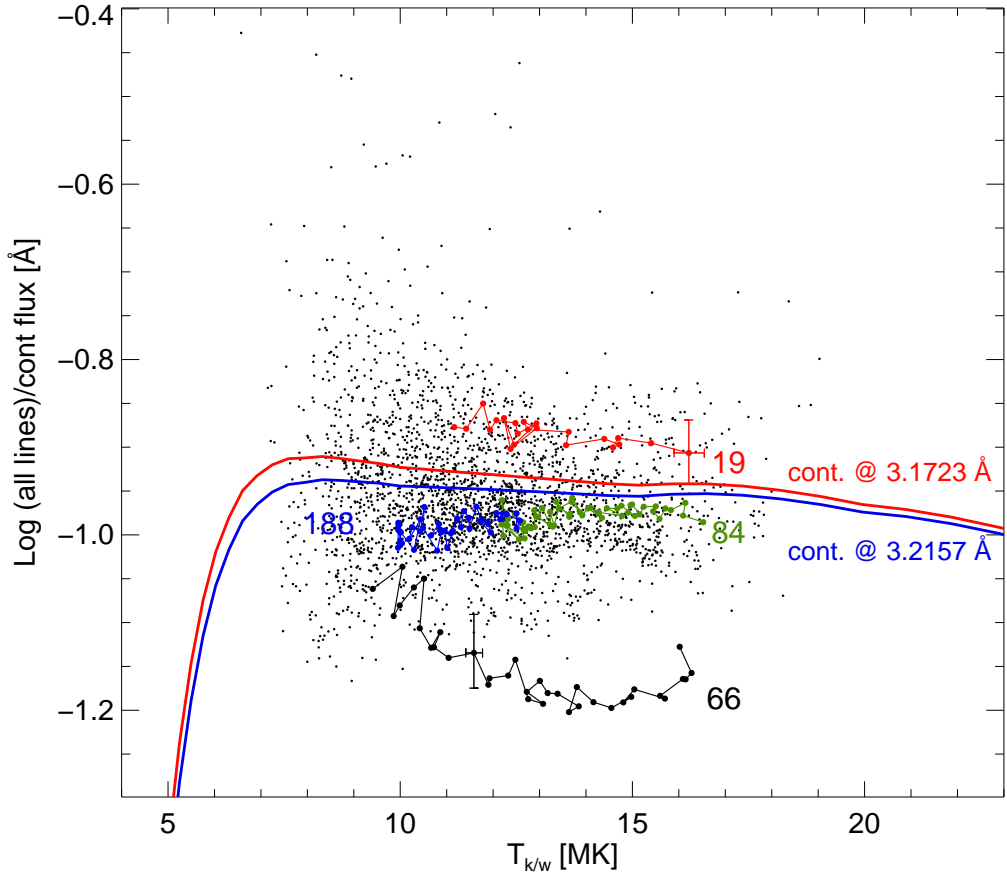}}
\caption{Logarithm of the intensity ratio of all lines in the BCS channel~1 range to the continuum plotted against temperature ($T_{\rm k/w}$). The red and blue curves are for the continuum at 3.1723~\AA\ and 3.2157~\AA\ respectively and an assumed Ca abundance, $A({\rm Ca})$, of 6.76 (average from \cite{jsyl98}) and Ar abundance $A({\rm Ar}) = 6.47$. Observed line-to-continuum points for the 2806 subintervals are plotted as small dots. Four individual flare decays are indicated by curves in red (flare no.~19, 1980 June~29), black (66, 1984 May~19), green (84, 1985 April~24), and blue (188, 1989 August~15). For two individual points (red and black) error bars are indicated.
\label{fig:fig8} }
\end{figure*}

Analysis of all spectra in the 2806 subintervals during the 194 flare decays was performed using the above procedures, finding first the total line emission in channel~1 spectra to emission in the continuum defined to be the mean of 20 lowest spectral irradiance bins of 0.0002~\AA\ width in the 3.169~--~3.228~\AA\ range. To estimate the Ca abundances, $A({\rm Ca})$, for each subinterval we used the theoretical line-to-continuum curves in Figure~\ref{fig:fig8} which are for an assumed Ca abundance $A({\rm Ca}) = 6.76$ and temperature $T_{\rm k/w}$. Departures of individual points from each curve give the value of $A({\rm Ca})$ with uncertainty from the statistical uncertainties in the estimations of the line and continuum photon counts and any time variations. Figure~\ref{fig:fig8} shows (as small black points) the logarithm of the line-to-continuum ratio plotted against the temperature $T_{\rm k/w}$ measured in each subinterval from the $k/w$ line ratio, using Voigt fits to each of the two lines (see Figure~\ref{fig:fig5}).  The solid colored curves in Figure~\ref{fig:fig8} are the line-to-continuum ratios determined from theoretical spectra that were calculated by \cite{phi18}.

%

In Figure~\ref{fig:fig8}, four individual flares are indicated by colored dots and connecting lines, with identifications shown in the figure caption (Table~\ref{tab:list-of-flares}). The time evolution of measured line-to-continuum ratios follows the temperature dependence of the theory curves approximately for three cases, indicating a nearly constant Ca abundance during the decay, although this is not the case for the flare 66 (1984 May~19).

\begin{figure*}
\centerline{\includegraphics[width=0.99\textwidth,clip=]{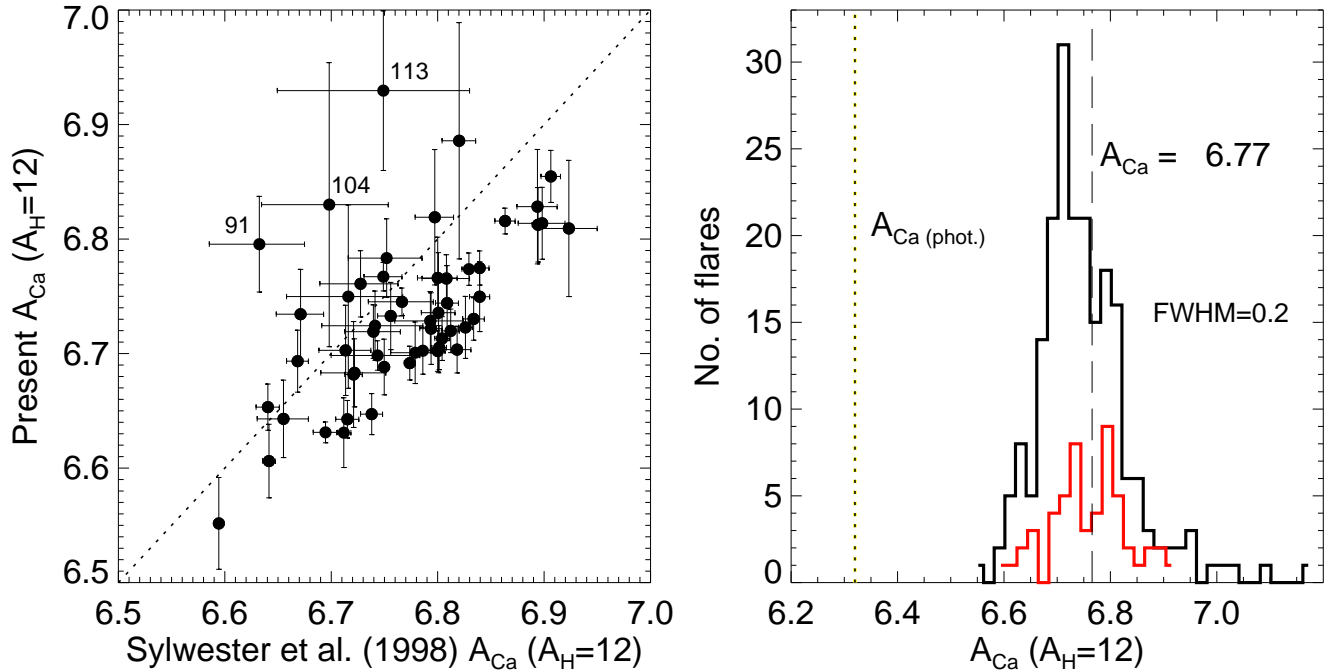}}
\caption{Left panel: estimates of the calcium abundance $A({\rm Ca})$ from the present  analysis  plotted against $A({\rm Ca})$ from \cite{jsyl98} for the common flare set.  Right panel: Distribution of calcium abundance $A({\rm Ca})$ averaged estimates for flares in this analysis (black histogram) and those in the \cite{jsyl98} analysis (red histogram). The average $A({\rm Ca})$ is 6.77, almost exactly the same as in the \cite{jsyl98} analysis. The photospheric Ca abundance (from \cite{lod21}), $6.32 \pm 0.03$, is indicated by the vertical dashed line.
\label{fig:fig9} }
\end{figure*}

Figure~\ref{fig:fig9} (left panel) is a plot of $A({\rm Ca})$ from the present analysis against the corresponding value from \cite{jsyl98} for the 53 flares in common. The three outliers (flare numbers marked 91, 104, 113) are considered not so reliable either because of a small number of subintervals (91, 104) or because of relatively fast temperature changes (113).

Figure~\ref{fig:fig9} (right panel) shows the distribution of $A({\rm Ca})$ values averaged over the decay of each flare (black histogram). The average for this distribution is 6.77 and the width (FWHM) is 0.20. Considered as a gaussian distribution, this FWHM corresponds to a standard deviation (s.d.) of 0.08. This uncertainty includes not only statistical uncertainties but any time variations over the flare decay; a typical uncertainty for individual spectra and individual flares (see Table~\ref{tab:list-of-flares}) is generally less than 0.08. The distribution of $A({\rm Ca})$ values for the \cite{jsyl98} flares is shown in red. The averaged $A({\rm Ca})$ is thus coincidentally nearly equal to that obtained by \cite{jsyl98}, $6.76 \pm 0.08$. The photospheric Ca abundance, from \cite{lod21}, is $6.32 \pm 0.03$ (marked by the dashed yellow vertical line) and is thus a factor 2.75 less than the estimate here.

\begin{figure}
\centerline{\includegraphics[width=0.75\textwidth,clip=]{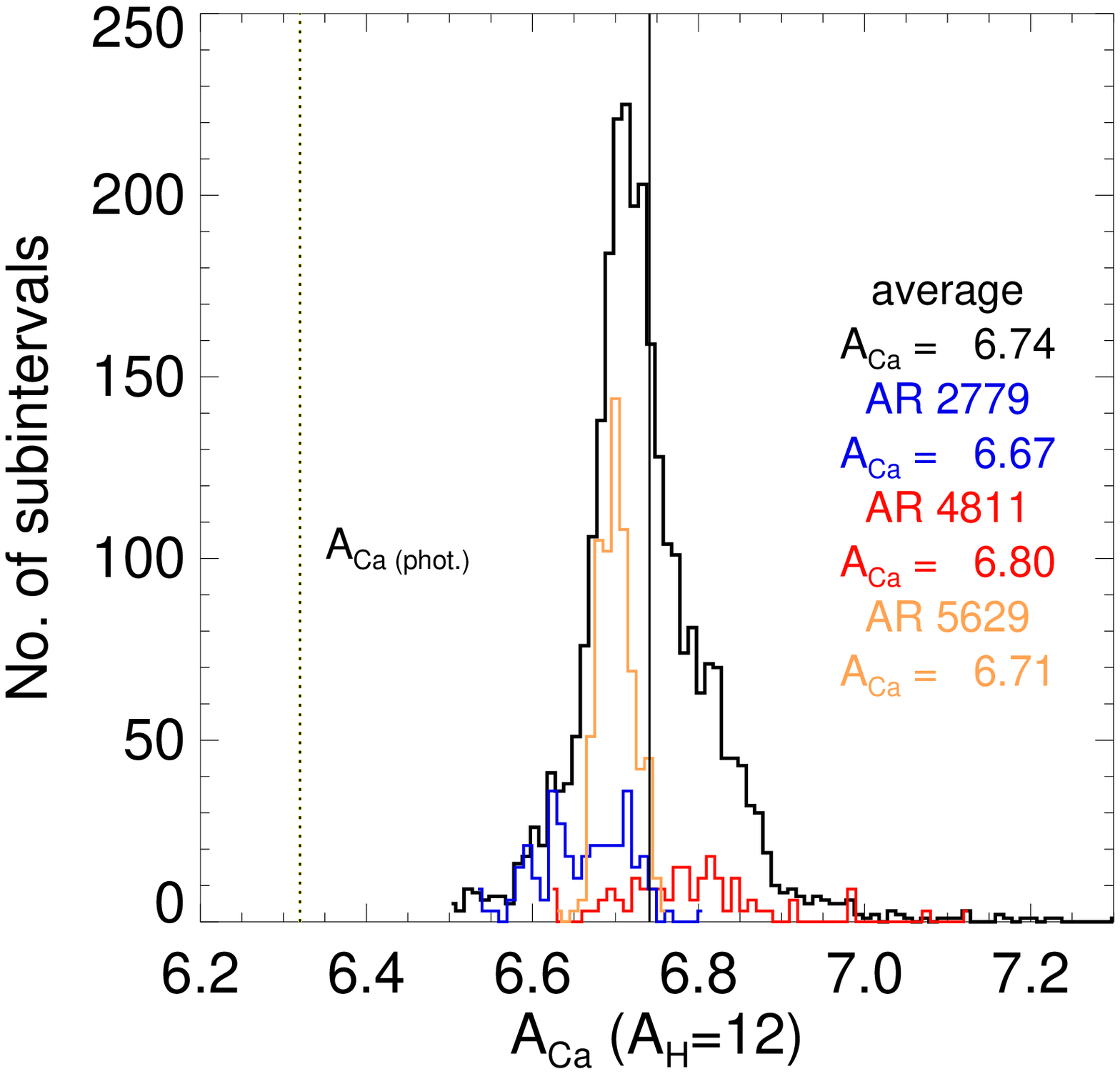} }
\caption{Distribution of estimated values of the calcium abundance $A({\rm Ca})$ for all 2806 subintervals (black histogram) and for subintervals of flares occurring in individual active regions (color-coded) -- the histograms are increased by a factor of 3 in this case to show them more clearly. The average $A({\rm Ca})$ for all subintervals is 6.74 (so slightly smaller than for averages over each of the 194 flares). The photospheric $A({\rm Ca})$ \citep{lod21} is indicated by the dashed vertical line.
\label{fig:fig10} }
\end{figure}

Since for some flares the line-to-continuum ratio varied significantly over the flare decay, a better estimate of the averaged $A({\rm Ca})$ is from the 2806 subintervals. The distribution is given in Figure~\ref{fig:fig10}. The mean value of $A({\rm Ca})$ for all subintervals is 6.74, slightly less than the average of flare decays (6.77). Several persistent active regions in the BCS lifetime produced many flares from which averaged $A({\rm Ca})$ values can be estimated; these are given in the figure. Thus $A({\rm Ca})$ is significantly less for active region AR~2779 (1980 November) than for AR~4811 (1987 May). The distribution for AR~5629 (1989 August) is much sharper than either of the other two active regions, with $A({\rm Ca})$ centered on 6.70. There is thus a suggestion of individual active regions having characteristic values of the Ca abundance, as was indicated by \cite{jsyl98} (Figure~4) with different distributions.

%

%

\begin{figure}
\gridline{\fig{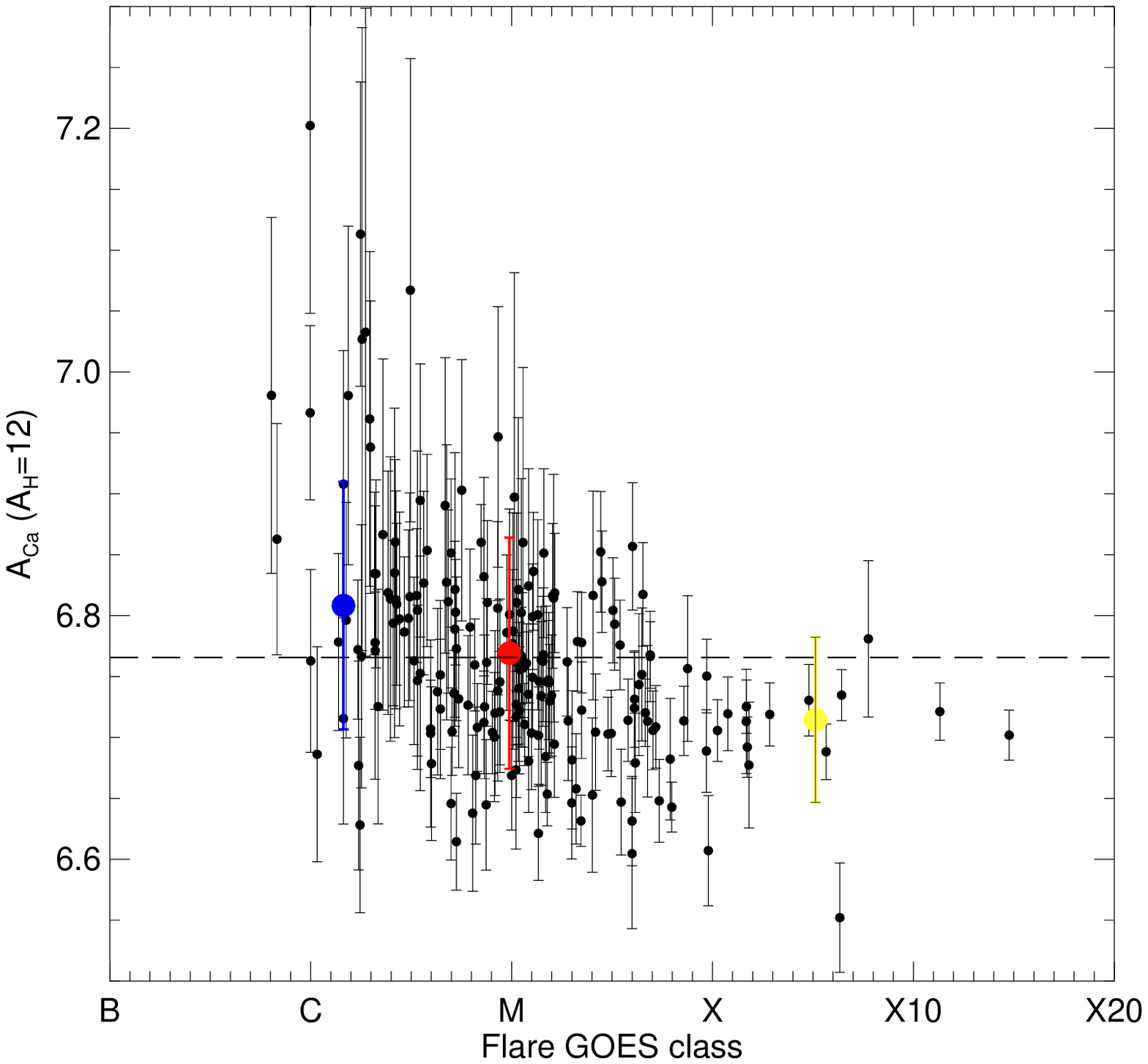}{0.45\textwidth}{(a)}
  \fig{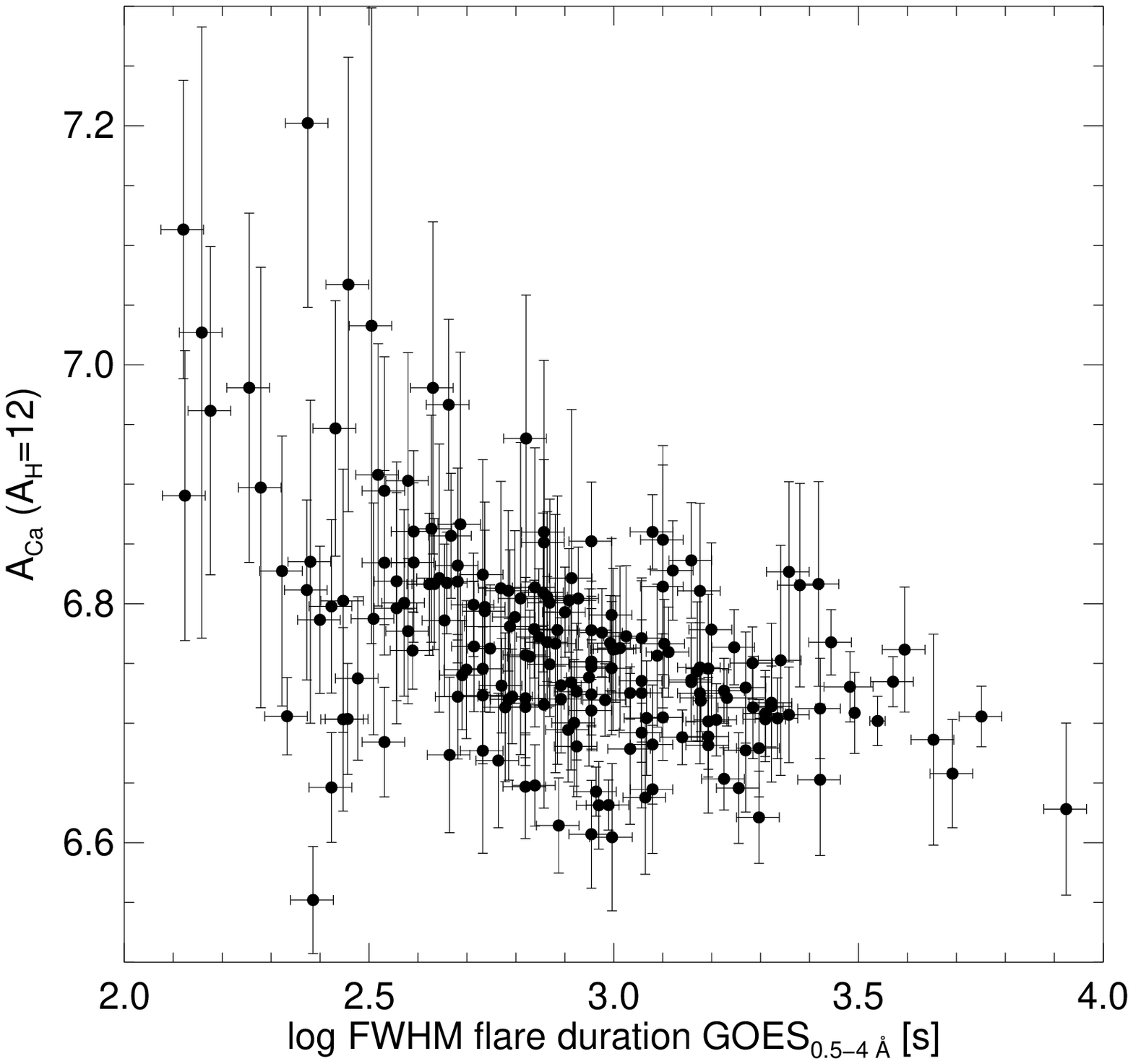}{0.45\textwidth}{(b)} }
\caption{Left panel $a$: Plot of the averaged $A({\rm Ca})$ for each of the 194 flares in this analysis against GOES X-ray class. The blue dot is average over class C, red dot over class M, yellow dot over class X. The dashed line is the average $A({\rm Ca})$ over the 194 flare decays. Right panel $b$: Plot of the averaged $A({\rm Ca})$ for each of the 194 flares in this analysis against total flare duration as measured from GOES 0.5~--~4~\AA.
\label{fig:fig11} }
\end{figure}

The Ca abundance variations between different active regions indicated by Figure~\ref{fig:fig8} suggest the possibility of long-term variations, similar to that found by \cite{bro17} using 1996~--~2006 data from the Extreme-ultraviolet Variability Experiment  (EVE) on Solar Dynamics Observatory. However, we found little or no correlation with either relative sunspot number (from the World Data Center WDC-SILSO, Brussels) or the solar radio flux at 10.7~cm (2800~Hz). We also investigated any correlation of $A({\rm Ca})$ with flare latitude, but found very little apart from a slight suggestion of a maximum for latitudes of about 24$^\circ$ and  a minimum at latitudes of about 10$^\circ$. A weak correlation of $A({\rm Ca})$ with flare X-ray class is, however, suggested by the plot in  Figure~\ref{fig:fig11} (left panel $a$). Here estimates of $A({\rm Ca})$ averaged over each flare decay are plotted against the flare GOES class. There is slight evidence for higher values of $A({\rm Ca})$ for flares with smaller GOES peak emission. The dashed line is $A({\rm Ca}) = 6.76$, the average over the 194 flare decays. The colored (blue, red, and yellow) dots are averages over classes C (B5 to C4), M (C5 to M4) and X (M5 or more) given together with corresponding (one standard deviation) dispersions. Rather more clearly is a slight correlation of $A({\rm Ca})$ with flare duration as shown in Figure~\ref{fig:fig11} (right panel $b$). The flare duration (FWHM) was determined from the GOES $0.5 - 4$~\AA\ light curve. In this case, there is a fairly marked decrease of $A({\rm Ca})$ from a value of up to 7.2 (short-duration flares) to about 6.7 (long-duration flares).

%

\section{Summary and Conclusions}\label{sec:concl}

The abundance of calcium has been investigated in this study from the decays of 194 flares seen with channel~1 of the SMM Bent Crystal Spectrometer and follows up the work of \cite{jsyl98}. There are significant improvements in the analysis techniques. Firstly, account is taken of crystal curvature deformations which affect line intensities and in particular temperature determinations from the $k/w$ line ratio. Secondly, we used the total calcium line emission in channel~1 rather than only the \ion{Ca}{19} resonance line, greatly enhancing the photon count statistics, and allowing improved abundance determinations through the fact that the temperature dependence is reduced. Finally, improved atomic data for the lines and continuum \citep{phi18} are applied.  A total of 194 flare decays (141 of which are new in this study) in the period 1980 to 1989 were included here. This encompassed the maximum of the very active Cycle 21, through the 1985/1986 minimum, to the rise of Cycle 22.

The resulting distribution of Ca abundances in Figure~\ref{fig:fig9} (right panel) to Figure~\ref{fig:fig11} clearly shows a wide range, with $A({\rm Ca}) = 6.77$ and the FWHM of distribution equal to 0.20 averaged over 194 flare decays or 6.74 (FWHM width 0.21) averaged over 2806 subintervals. Values of $A({\rm Ca})$ (Figure~\ref{fig:fig9}) range from 6.6 to 7.2. There appears to be no significant correlation with solar activity indices (sunspot number and F10.7 radio emission) or with flare latitude, but there is a weak correlation of $A({\rm Ca})$ with GOES flare class (larger Ca abundances for relatively weak flares) and with flare duration as defined by the GOES (0.5~--~4\AA) light curve (larger $A({\rm Ca})$ for shorter-duration flares: see Figure~\ref{fig:fig11}).

Our $A({\rm Ca})$ values are larger by a factor of 2.6 than those found from photospheric and meteoritic studies $A({\rm Ca})= 6.32 \pm 0.03$ \citep{lod21}, with a total range of approximately 1.9 to 7.6 times photospheric or meteoritic. Calcium, then, behaves like a low-FIP element in the manner described by \cite{fel92b} although the analysis of \cite{jsyl98} is confirmed that there is significant flare-to-flare variability.

The widely cited theory of \cite{lam04,lam12} ascribes the enhancement of low-FIP elements in the solar corona and solar wind to a ponderomotive force associated with Alfv\'{e}n or fast-mode magnetohydrodynamic waves traveling upwards or downwards. Recent updates by \cite{lam21} include calculations of the changes in coronal element abundances for a combination of the ratio of magnetic fields in the corona to the photosphere and fast-mode wave amplitudes at the plasma beta $\beta = 1$ level. These were intended for comparison with the X-ray flare observations of \cite{kat20} and \cite{den15}, but here we apply them to our results.

Our Ca abundance enhancements over photospheric, 2.6 with a range of 1.9 to 7.6, are compatible with the calculated values for a range of photospheric-to-coronal field strength $B_{\rm cor} / B_{\rm phot}$ ratios and relatively small ranges of fast-mode wave amplitudes $v_{fm}$. Thus, for $B_{\rm cor} / B_{\rm phot} = 0.5$, $v_{fm}$ in a range of 0 to 10~km~s$^{-1}$ would explain our observed Ca enhancements over photospheric. This range of parameters leads to enhancements or diminutions for other abundant elements, including Si, S, Ar, and Fe. The adopted abundances in Table~\ref{tab:abundances} have values that are, relative to photospheric, Si (0.5), S (0.5), Ar (1.2), Fe (2.8). The enhancements for Ar, Ca, and Fe can be reconciled with $B_{\rm cor} / B_{\rm phot} = 0.5$ and $v_{fm}$ a little less than 8~km~s$^{-1}$. However, this then disagrees with determinations of sulfur and silicon flare abundances that are lower than or comparable to photospheric: the \cite{lam21} theory predicts enhancements for these elements. The most recent Si abundance determinations listed in Table~\ref{tab:abundances} indicate that $A({\rm Si})$  is similar to or slightly less than photospheric, from sources which are fairly secure: RESIK \citep{bsyl15} and OSO-8 \citep{vec81} crystal spectrometer results based on line-to-continuum ratios (instrumental background due to crystal fluorescence being negligible or accurately assessed) and the relatively high-resolution broad-band Suzaku XIS flare spectra of \cite{kat20}. This also applies to the sulfur abundance determination which is appreciably less than photospheric, again based on the RESIK, OSO-8, and Suzaku results. Thus, there is some incompatibility with the recent theoretical calculations of \cite{lam21} that may require some adjustment of the model. Otherwise the abundances of iron and argon are in good agreement with the \cite{lam21} calculations.

%
\begin{acknowledgments}

We thank Professor Christopher Rapley for his continued support of this work and his knowledge of the Bent Crystal Spectrometer. We acknowledge financial support from the Polish National Science Centre grant number UMO-2017/25/B/ST9/01821. The {\sc chianti} atomic database and code is a collaborative project involving George Mason University, University of Michigan (USA), and University of Cambridge (UK).

\end{acknowledgments}

\facilities{Solar Maximum Mission (BCS)}
\software{SolarSoft Interactive Data Language \citep{fre98}, {\sc chianti} \citep{delz15}}

\newpage


%
\bibliography{RESIK}
\bibliographystyle{aasjournal}
%




\end{document}